\documentclass[acmsmall, nonacm]{acmart}
\usepackage{xspace}

\AtBeginDocument{%
  }

\setcopyright{acmlicensed}
\copyrightyear{2025}
\acmYear{2025}
\acmDOI{10.1145/3746648}

\acmJournal{TORS}
\acmVolume{1}
\acmNumber{1}
\acmArticle{1}
\acmMonth{1}

\usepackage{todonotes}
\presetkeys%
    {todonotes}%
    {inline,backgroundcolor=yellow}{}

\begin{document}

\newcommand{\rsforgood}{RS4Good\xspace}
\newcommand{\mt}[1]{\textcolor{blue}{[MT] #1}}
\newcommand{\as}[1]{\textcolor{green}{[AS] #1}}

\title{Recommender Systems for Good (\rsforgood): Survey of Use Cases and a Call to Action for Research that Matters}
\subtitle{Opinion Paper}

\author{Dietmar Jannach}
\email{dietmar.jannach@aau.at}
\orcid{0000-0002-4698-8507}
\affiliation{%
  \institution{University of Klagenfurt}
  \country{Austria}
}

\author{Alan Said}
\email{alansaid@acm.org}
\orcid{0000-0002-2929-0529}
\affiliation{%
  \institution{University of Gothenburg}
  \country{Sweden}}

\author{Marko Tkalčič}
\email{marko.tkalcic@gmail.com }
\orcid{0000-0002-0831-5512}
\affiliation{%
  \institution{University of Primorska}
  \country{Slovenia}}

\author{Markus Zanker}
\email{markus.zanker@acm.org}
\orcid{0000-0002-4805-5516}
\affiliation{%
  \institution{Free University of Bozen-Bolzano and University of Klagenfurt}
  \country{Italy and Austria }}

\renewcommand{\shortauthors}{Jannach et al.}

\begin{abstract}
In the area of recommender systems, the vast majority of research efforts is spent on developing increasingly sophisticated recommendation models, also using increasingly more computational resources. Unfortunately, most of these research efforts target a very small set of application domains, mostly e-commerce and media recommendation. Furthermore, many of these models are never evaluated with users, let alone put into practice. The scientific, economic and societal value of much of these efforts by scholars therefore remains largely unclear. To achieve a stronger positive impact resulting from these efforts, we posit that we as a research community should more often address use cases where recommender systems contribute to \emph{societal good} (\emph{\rsforgood}). In this opinion piece, we first discuss a number of examples where the use of recommender systems for problems of societal concern has been successfully explored in the literature. We then proceed by outlining a paradigmatic shift that is needed to  conduct successful \rsforgood research, where the key ingredients are interdisciplinary collaborations and longitudinal evaluation approaches with humans in the loop.\footnote{Paper accepted for publication in ACM Transactions on Recommender Systems}
\end{abstract}

\ccsdesc[500]{Information systems~Recommender systems}

\keywords{Recommender Systems, Evaluation, Applications}

\maketitle

\section{Introduction}
\label{sec:introduction}
Whenever you read a paper on recommender systems published in the past twenty-five years, chances are good that it has several or all of the following characteristics: (i) the paper proposes a new recommendation algorithm, (ii) the algorithm is evaluated in the movie or e-commerce domain, and (iii), the evaluation is focusing on prediction accuracy and it is based on an offline experiment without involving humans in the loop~\cite{JannachZankerEtAl2012}. As a result, we nowadays have hundreds---if not thousands---of algorithmic proposals and machine learning (ML) models at our hands that were designed for, or at least evaluated, on some movie or e-commerce dataset. Yet these achievements are seemingly not enough. Despite or due to all this progress, research addressing similar problems flourishes and, nowadays, involves increasingly complex deep learning architectures and, most recently, large language models with an immense carbon footprint~\cite{Vente2024fromClicks,Spillo2024Towards}.

In terms of application domain, there is of course no doubt that recommendations can create substantial business value in e-commerce environments, see~\cite{jannachjugovactmis2019}. There are also various papers from large players in the media streaming market reporting on the benefits of providing personalized movie or video recommendations, including Netflix
and YouTube~\cite{Steck2021DeepLearning,Gomez-Uribe:2015:NRS:2869770.2843948,Covington:2016:DNN:2959100.2959190}. However, due to the strong focus on offline experimentation in scholarly research, we unfortunately do not possess a lot of evidence on the practical effectiveness of the countless algorithmic proposals mentioned above. Will all these algorithmic contributions with their mostly small increases in offline prediction accuracy lead to more user engagement, higher retention rates, higher sales and more satisfied users? For many algorithmic proposals, the answer is probably 'no'. Netflix actually never implemented the winning solution of the million-dollar Netflix Prize competition in production\footnote{\url{https://netflixtechblog.com/netflix-recommendations-beyond-the-5-stars-part-1-55838468f429}}, and in a later paper~\cite{Gomez-Uribe:2015:NRS:2869770.2843948} made a rather disappointed statement about the often limited correlation between outcomes of offline experiments and the success of a deployed model.

All that puts us as researchers in a situation where we have developed countless recommendation algorithms over the years, of which however only a very tiny fraction has been tested `in the wild'. Even worse, there is both evidence from the academic literature and from industry reports that gains in offline accuracy do \emph{not} reliably translate to increased consumer or business value in practice. Nonetheless, despite all these known limitations we observe a constant and maybe even increasing inflow of new algorithmic papers that are evaluated in offline experiments on datasets from a limited set of domains.

For us as a recommender systems research community, this raises the following question: why do we continue to focus so narrowly on a few application domains and methodologies? Actually, the literature contains a multitude of examples where recommender systems are applied to a variety of application domains beyond movies and e-commerce. Moreover, the value that is created in several of these settings goes beyond the creation of economic value for the platform provider as well as beyond individual hedonic or utilitarian needs of consumers. We may, for instance, think of recommender systems that guide users to a healthier life style, recommender systems that stimulate energy-saving behavior, or systems that help users achieve self-actualization goals.

Given that these application areas are largely underexplored in the literature, we call the research community to re-focus our combined efforts to areas where recommender systems can contribute to the achievement of \emph{societal} goals. Following ideas of recent initiatives like `AI4Good'\footnote{https://ai4good.org/} or `IR4Good'\footnote{\url{https://www.ecir2024.org/2023/07/24/call-for-ir-for-good-papers/}}, we envision the development of an `\rsforgood' movement in the recommender systems community. As a result, we hope that the future research efforts in this area will be better aligned with the pressing challenges that we face as a society. However, as we will lay out in more depth in the following sections, to achieve these goals it will not be sufficient to only use different datasets, but a paradigmatic shift is required, instead. It involves the following components: (i) a focus on problems of societal relevance, (ii) a multi-disciplinary research approach involving humans in the loop, and, (iii) the consideration of the longitudinal effects of recommender systems.

The rest of this paper is organized as follows. Next, in Section~\ref{sec:topics}, we will explain why \rsforgood ambitions must go beyond recent initiatives to only avoiding harm, and we will outline a number of \rsforgood application areas. Section~\ref{sec:methodology} then describes the implications of a shift to \rsforgood use cases in terms of the applied research methodology.

\section{\rsforgood: Topic Areas}
\label{sec:topics}
In this section, we position the \rsforgood initiative with respect to recent developments in the area of `responsible recommendation' and then review a number of potential application areas for future \rsforgood developments.

\subsection{Defining \rsforgood: From Avoiding Harm to Doing Good}
Recommender systems are commonly designed to create positive effects and different types of value for the involved stakeholders. Typical examples of such values include the avoidance of information overload for end users or increased sales for businesses. During the past few years, it however became more and more evident that recommender systems may also lead to harm and certain undesired negative effects. In 2011, the discussion of possible `filter bubbles' that may emerge from personalized recommendation led to notable public attention~\cite{Pariser2011Bubble}. Since then, we observed increased awareness of potential problems of recommender systems, leading to various corresponding research works and initiatives which may be collectively placed under the umbrella terms `responsible recommendation' or `trustworthy recommendation'.

The main aspects in this area are the awareness of recommendation biases and the consideration of goals such as fairness, privacy protection, interpretability and explainability, various forms of trustworthiness, robustness, auditability, or compliance with laws and regulations. Beyond the issues of `filter bubbles' and `echo chambers', central topics in the news and media domain also include the avoidance of the spread of misinformation, polarization, and radicalization. Corresponding surveys on these topics can be found in~\cite{Wang2024Trustworthy,Chen2023Bias,dejdjoo2023fairness,ekstrand2021fairness,Wang2023SurveyFairness,Deldjoo2021SurveyAdversarial,Si2020Shilling,Zhang2020Explainable,elahi2021aiethics}.

The central underlying motivation of many of these approaches, as mentioned, is rather to avoid harm than to \emph{proactively} target at creating a positive value for society. Admittedly, these aspects may often appear related and overlapping. In our view, however, these approaches set apart the \emph{underlying goals} of the recommender system.

Let us consider the example of a content recommender system on social media. Such recommender systems are often designed to maximize user engagement, with the ultimate goal of increasing user retention, ad revenue, and company growth. To maximize the engagement, a recommendation algorithm might now learn that promoting controversial or even hateful content leads to a high number of interactions, e.g., in terms of user comments. While this indeed may have positive short-term effects on the mentioned business metrics, it may have negative effects as well, e.g., on the public reputation of the platform. As a consequence, the platform might implement certain measures in the recommendation algorithm to delimit the promotion of potentially harmful or problematic content to a certain extent.

Such avoidance or mitigation strategies are different from our notion of \rsforgood that we are advocating in this paper. Rather than to avoid problems that may for example be caused by the underlying business goals or by the unintended reinforcement of biases in the underlying data, social benefit is part of the `genes' and main purpose of an \rsforgood recommender system. Certainly, this does not contradict the assumption that recommender systems are designed to create economic value for providers. Let us think of an alternative social media platform that tries to proactively balance the coverage of content from different political viewpoints, thereby implementing a corporate mission to reduce radicalization and polarization in society. However, such a societal mission does not necessarily mean that economic goals may not be pursued in parallel. The provision of balanced and diversified content may in fact be a key factor of positioning the platform on the market, similar to independent publishers or traditional newspapers in media environments.

\subsection{Selected Use Cases of \rsforgood Research}
In their recent work, Felfernig et al.~\cite{felfernig2024sustainable} discussed the potential role of recommender systems to support the United Nation's 17 Sustainability Development Goals (SDGs). These SDGs comprise both areas that are well aligned with typical goals of recommender systems (e.g., `economic growth') as well as areas which, as of today, are barely related to existing recommender systems research (e.g., `life below water'). In our following discussion of the \rsforgood research landscape, we will therefore not follow the UN's SDG organization, but rather structure our work along application domains where the use of recommendation systems has been already explored at least to a certain extent. We emphasize that our discussion and categorization is not meant to be exhaustive. Instead, it shall serve as an illustration and motivation for the many possible directions that the research community should explore in more depth in the future.

\begin{figure} [ht]
  \centering
  \includegraphics[width=0.8\textwidth]{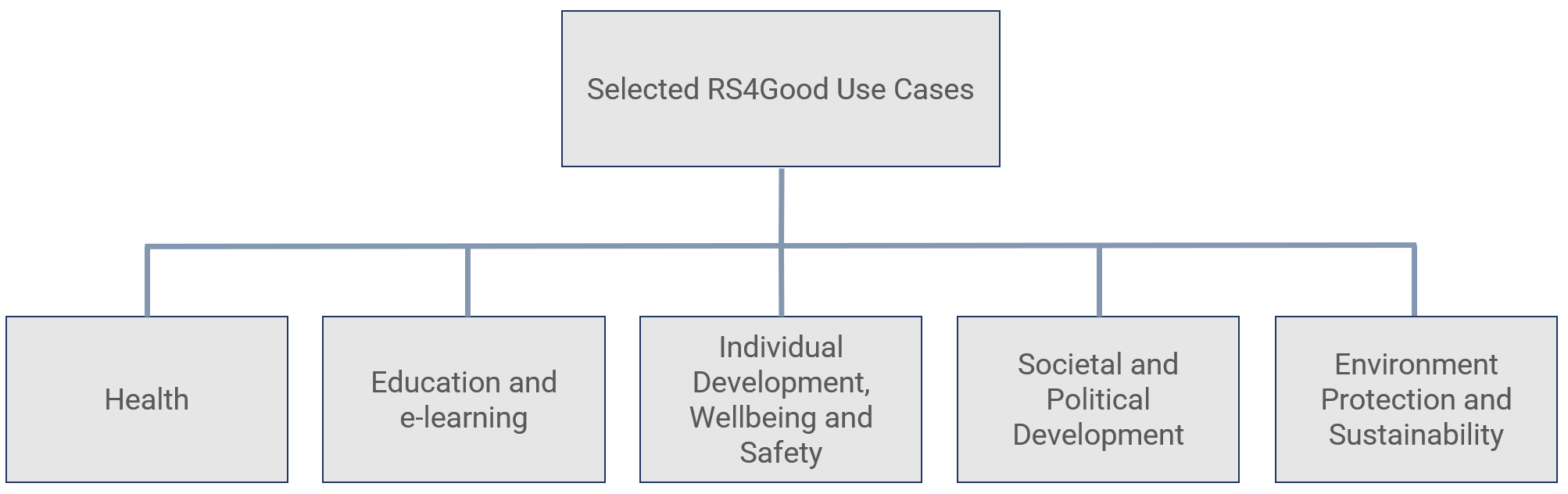}
  \caption{Selected \rsforgood Use Cases}\label{fig:use-cases}
\end{figure}

We note that our discussion of existing works is mainly organized in terms of application domains for presentation purposes. We however recall that whether a recommender systems is considered to be `for good' does not necessarily depend on the application domain. A recommendation system in the e-commerce domain, for example, can be guided purely by economic objectives~\cite{ZankerBricmanEtAl2006}. It may however also have elements of an \rsforgood system, e.g., when the recommender is designed to prioritize sustainable product alternatives. As mentioned in the previous section, it depends on the underlying goals that are implemented in the recommender system. In the following, we review a number of \rsforgood systems in different application contexts. Figure~\ref{fig:use-cases} illustrates the main topic areas.\footnote{We deliberately leave out the general goal of \emph{economic growth} in this list while acknowledging that the growth of individual businesses and the economy may generally contribute to positive societal developments. Other categorizations of existing works are possible as well, e.g., by distinguishing between individual and societal goals. We discuss questions of goal identification and goal setting later in Section~\ref{subsection:challenges-barriers-rs4good}.}

We re-iterate that our discussions of individual research works in each application area are not intended to represent a comprehensive review of the partially quite rich literature, e.g, in the educational domain, but serve as motivational examples. An overview of selected in-depth surveys in relevant domains can be found in Table~\ref{tab:surveys}.

\begin{table}[h!]
\caption{Selected survey works in different domains.}
\label{tab:surveys}
\begin{tabular}{ll}
   \emph{Domain} &  \emph{References} \\ \toprule
   Health and Healthcare &  \cite{DeCroon2021HealthSurvey}, \cite{Etemadi2023Systematic}, \cite{Pincay2019HRS}, \cite{Tran2021} \\
   Education and e-learning & \cite{Bodily2017studentfacing}, \cite{Silva2023EduSurvey}, \cite{Milicevic2014elearning}, \cite{Urdaneta2021education} \\
   Individual Development \& Well-Being &   \cite{Elsweiler2022}, \cite{Felfernig2024Sports}, \cite{Mahajan20215systematicfood}, \cite{Tran2018HealthyFood} \\
   Societal and Political Development  & \cite{Bauer2024Whereare}, \cite{Kamal2024HigherEdRS} \\
   Environment Protection \& Sustainability & \cite{felfernig2024sustainable}, \cite{zhou2024advancingsustainabilityrecommendersystems}
\end{tabular}
\end{table}

\subsubsection{Health}
The health domain is a prime example where recommender systems can be applied for societal good in various ways. One notable survey of recommender systems in healthcare can be found in~\cite{Tran2021}. In this survey, the authors identify the following main application areas.
\begin{itemize}
\item \emph{Food recommendation:} Use cases involve the recommendation of proper diets, better nutrition, or healthier food alternatives.
\item \emph{Drug recommendation:} Drug recommendation systems are designed to help patients and health-care professionals to identify medications for certain patient conditions or to predict drug side-effects.
\item \emph{Health status prediction:} Recommendation techniques like collaborative filtering can be leveraged to make better predictions of a patient's risk factors given their current health status.
\item \emph{Physical activity recommendation:} Approaches in this area were for example developed to suggest exercises or workouts to patients, e.g., to achieve their calorie burn goals or to perform their activities to improve sleep~\cite{Upadhyay2020sleep}.
\item \emph{Healthcare professional recommendation:} Recommender systems have also been developed to help patients find doctors with the best expertise for their health-related issues.
\end{itemize}

A number of works in the literature cover additional topics that go beyond this classification, e.g., the recommendation of messages within a smoking cessation app~\cite{HorsFraile2016smoking}. Such works fall into the more general category of \emph{behavior change support systems}~\cite{Oinas2010behavior}. Some of the works in the literature are also targeted at certain groups of people or patients, for instance, elderly~\cite{Espin2014elderly,Herpich2014TowardsGamified,Hammer2014Designof} or people with autism~\cite{MAURO2022autism}, dementia~\cite{Steinert2022dementia}, or social anxiety~\cite{Ameko2020Offline}.
Today, many of the more recent health-related recommendation approaches are implemented as (smartphone) apps. This leads to additional research questions beyond those related to the core recommendation task. The \emph{credibility} of a health-related app is for example, important for the adoption of the recommendation, as analyzed in~\cite{Oyibo2018credibility}. Also there are often \emph{ethical} aspects to consider, in particular in the context of mental health apps~\cite{valentine_recommender_2023,Rohani2020Mubs}. Of course, it is often highly important to limit the risks of inadequate recommendations in such a critical domain. This can, for instance, be achieved by providing adequate explanations alongside recommendations, and by leaving the final decision of the chosen health intervention to the medical expert~\cite{deLima2021}.

Generally, health recommender systems is an active research area, as also evidenced by a series of health-related workshops held in conjunction with the ACM RecSys conference\footnote{\url{https://healthrecsys.github.io/}}. The literature on the topic is however quite scattered across various publication outlets, see also the survey on recommenders for digital health in~\cite{Cheung2019HowRecommender}. Various works in this area are published in computer science outlets, others in medical journals, indicating the potential for more interdisciplinary research.

\subsubsection{Educational Recommender Systems and e-Learning}
Another quite active and established application domain for \rsforgood is the use of recommendation technology to support human/student \emph{learning} processes, e.g., in the form of course recommendations or learning paths suggestions in e-learning environments. Works in this area fall into the broader category of `AI in education', which is one of UNESCO's focus topics\footnote{\url{https://www.unesco.org/en/digital-education/artificial-intelligence}}. Also, e-learning recommendations represent one form of \emph{Technology-enhanced Learning} (TEL)~\cite{Duval2017TEL}.

Several types of items can be recommended in e-learning recommender systems, including courses and course paths, college majors, papers, web pages, lessons and various other learning resources. Various types of traditional recommendation algorithms---both content-based, collaborative filtering and hybrid ones---have been explored in the literature~\cite{Milicevic2014elearning}.
In some approaches, the recommendation of learning activities (and materials) is based on predicting the student's performance on a task that she or he has not solved yet. Given these predictions, the set of recommendations is then determined based on the e-learning system's specific goals. For example, the system could recommend materials that the student assumedly will like and will most probably be able to master. Following the exact opposite strategy, the system may however also recommend materials where it assumes gaps in the student's knowledge. In this context, it is important to note that recommending only courses and materials for which the system assumes that the student will receive a good grade may actually be counterproductive, e.g., when the system starts recommending only easy courses. Instead, one could argue that recommending more challenging courses---or courses that lead to a well-rounded overall educational profile---will contribute better to a student's growth and development.

An in-depth discussion of the \emph{challenges} of building an e-learning recommender systems can be found in~\cite{Milicevic2014elearning}. A recommender system not only has to understand the learning goals and the prior knowledge of the student, it has also to take into account personal characteristics, such as the student's learning strategy. Moreover, recommendations must be made at the right point in time and within an appropriate interactive tool environment. Generally, technology-enhanced learning processes and educational recommender systems are a multi-faceted topic, and research in this direction usually requires an interdisciplinary approach~\cite{Balacheff2009}. Furthermore, the evaluation of e-learning recommender systems remains difficult. As mentioned above, depending on the system's goal and strategy, one or the other type of learning content may constitute appropriate recommendations for given students and their assumed educational goals. Furthermore, the \emph{true success} of the system, i.e., if the student's knowledge or learning process, was improved, can usually only be evaluated from a longitudinal perspective.

\subsubsection{Individual Development, Wellbeing and Safety}
Besides healthcare and education, a number of works address alternative ways of how recommender systems can contribute to \emph{individual} development and wellbeing of users. Some of these works are related to personal health aspects, e.g., works by Elsweiler et al.~\cite{Elsweiler2017FoodRecipe}, who evaluate how recommender systems can \emph{nudge} users to choose healthier recipes such as dishes with more favorable nutritional characteristics on food-focused online sites. A related study on the role of nutrition labels on food choices can be found in~\cite{DBLP:conf/recsys/StarkeMT22}. A more general approach to support personal \emph{wellbeing} with recommender systems was proposed by Gryard et al.~in \cite{GYRARD2020100083}. Their work is motivated by the increased availability of sensors on Internet-of-Things wearable devices that can measure physiological signals. In their system, these sensor signals are fed into a rule-based reasoning engine that recommends actions to (re-)achieve `happiness', e.g., do physical activities, increase the intake of certain natural products, or apply procedures from alternative medicine. A related IoT-based approach for improved mental health was later proposed by the authors in~\cite{Gyrard2022Interdisciplinary}. Unfortunately, no evaluation with real users has been performed for these approaches yet.

In this context, we acknowledge that some of the works discussed here, i.e., those that target individual wellbeing, could also be seen to fall under the category of health recommender systems discussed above. In the individual wellbeing category, we consider works that do not focus primarily on the clinical aspects of health or contribute to health as a societal goal. Instead, the works discussed here are often about behavior change, prevention, or individual goal-setting. Consider, for example, the work by \citet{Radha2016Lifestyle}. In this research, the authors investigate a patient-centered recommender system that provides personalized lifestyle modification advice for blood pressure management, considering also the individual patient's ability level. Similarly, the user's ability level and task difficulties were  also taken into account in a nutrition recommender systems in~\cite{Schaefer2019Rasch} and in a behavior-change recommender system in~\cite{Torkamaan2022RecChallenges}.

Recreational sport is often considered to be another important contributor to wellbeing. The use of recommendation technology and machine learning techniques for amateur marathon runners was for example proposed in Smyth et al.~\cite{Smyth2022marathon}. The primary recommendation tasks in their approach range from the suggestion of personalized training plans to pacing strategies and the prediction of realistic finish-times. Secondary use cases for runners include the selection of running events, running routes, training partners, gear and equipment as well as advice regarding nutrition and recovery. In~\cite{Smyth2022marathon}, a number of these uses cases for recommendation technology are discussed in more depth. The evaluation of the proposed approaches was however largely limited to offline experiments so far.

Going beyond the realm of sports and health, Knijnenburg et al.~\cite{Knijnenburg2016selfactualization} propose to develop a new class of recommender systems for \emph{self-actualization}. Instead of just recommending the ``best'' items to users, such systems are designed to \emph{``support users in developing, exploring, and understanding their own unique tastes and preferences.''} As a consequence, future recommender systems for self-actualization should help users break out of filter bubbles and rather support users instead of replacing them in their decision-making tasks. A possible experiment to evaluate such a recommender system was later outlined in~\cite{Wilkinsontesting}. The results from a user study in the news recommendation domain are discussed in~\cite{Sullivan2019selfact}. In this study, it was found that the reading intention of participants is influenced by their stated self-actualization goals, which could either be \emph{Broaden Horizons} or \emph{Discover the Unexplored}.
In particular in this area of self-actualization and personal development, it generally seems highly important to properly understand the user's goals, and to actively involve them in the recommendation process instead of just interpreting their observed behavior~\cite{Ekstrand2015Behaviorism,Liang2024Enabling}.

As a final example in the context of individual development, wellbeing and safety we consider recommender technology contributing to personal safety. Specifically, a number of approaches have been proposed to recommend privacy settings to users in the online sphere. An early work~\cite{Ghazinour2013Monitoring} on recommending privacy settings on Facebook was based on categorizing users into different groups according to their assumed privacy values. A more elaborate technical approach to recommend image sharing settings was later proposed in~\cite{Jun2018Leveraging}. A particularity of this work is that the content itself, i.e., the images, are analyzed for content sensitiveness. Later on, the work in~\cite{Salem2021OharsPrivacy} aims at considering the user's disclosure preferences while protecting their privacy. A specific model for the domain of fitness devices can be found in~\cite{Sanchez2020privacyfitness}.

\subsubsection{Societal and Political Development}
A number of \rsforgood use cases can be identified that target at the societal or political development of a society as a whole. In the context of \emph{public opinion formation}, for instance, recommender systems are nowadays widely employed on news sites and social media channels. While news recommender systems can be valuable both for consumers and providers, they are often associated with potential problems such as the creation of filter bubbles or leading to polarization. \rsforgood initiatives, as indicated above, should however go beyond avoiding potential harm via \emph{responsible media recommendation}~\cite{elahi2021aiethics}. Instead, they may, for instance, aim at supporting a societal editorial mission or a given service mandate as many public European broadcasters do. In \cite{Clark2023BBC}, a discussion of challenges of building a recommender system for the British Broadcasting Corporation (BBC) can be found. Grün and Neufeld in~\cite{Gruen2023ZDF} highlight the importance of transparency, fairness and exploration goals for the public German media service ZDF. Related discussions can be found in~\cite{Smets2020Designing}, where authors advocate for participatory design practices and open data to address some of the challenges of digital public services.

Also going beyond the goal of merely avoiding harm, Lu et al.~\cite{Lu2020Beyond} conduct an online study in the news domain to assess how personalized recommendations influence the reading behavior of users in the context of editorial values, like diversity and coverage. Furthermore, an intervention study reveals that a re-ranking algorithm is effective to steer readers to a more \emph{dynamic} reading behavior without a loss of accuracy. In a related work, Diakopoulos et al.~\cite{diakopoulos2024leveragin} more generally discuss the importance of considering professional ethics when designing responsible AI-based solutions for journalism. Furthermore, Bauer et al. review the consideration of journalistic values\footnote{A recent, more general discussion of how to implement \emph{human values} in recommender systems can be found in~\cite{Stray2024HumanValues}.}  in news recommender systems~\cite{Bauer2024Whereare}.

Also in the realm of \emph{social} media, personalization and recommender systems are often assumed to be the cause of echo chambers, filter bubbles and polarization~\cite{Donkers2021Dual,Stray_2022_depolarize}. Donkers et al.~\cite{Donkers2021Dual} investigate the phenomenon of `dual' echo chambers with an agent-based modeling and simulation approach. Their study reveals that counteracting the two types of echo chambers requires different proactive diversification and depolarization strategy. In a similar vein, Stray~\cite{Stray_2022_depolarize} studies how to design recommender systems for depolarization. Diversification of the presented content is considered as one possible strategy that can work against polarization in some context, but it may even make things worse. According to~\cite{Stray_2022_depolarize}, recommenders should generally be designed to deprioritize content that is found to be polarizing, and polarization measures should be incorporated into systems and continuously monitored.
In that context, \citet{OvadyaThorburn2023} propose what they call \emph{bridging systems}: systems that promote mutual understanding and trust and which thereby help reducing polarization. Among other examples, the authors discuss how recommender systems on social media could be designed to implement a bridging functionality. Instead of designing the system to maximize engagement, which may be achieved by recommending controversial content, a bridging recommender should for example emphasize content which is endorsed by people who in the past had diverse viewpoints.

While the news and social media domains are key use cases for recommender systems, there are various other scenarios where recommender systems can support societal development. In the educational sector, for example, Wilson et al.~\cite{Wilson2009Smartchoice} propose an online recommender system that is particularly designed to help families of low socioeconomic status to select an appropriate school that matches parents' preferences and the students' needs and abilities. In the same educational realm, Milton et al.~\cite{Milton2019StoryTime} develop a book recommender system for kids that aims at increasing their reading interest and literacy.

A number of further research works aim at fostering societal development and political engagement through recommender systems in terms of supporting \emph{e-participation}~\cite{Cantador2017Personalized,DBLP:conf/recsys/Segura-TinocoC21} or \emph{e-governance}~\cite{Cortes-Cediel2017Recommender} processes. Furthermore, the authors of~\cite{Diaz-Agudo2021Towards} propose a recommender system to increase \emph{social cohesion} by enabling each member of the community to feel that they are part of the cultural heritage of society. Cultural content recommendations are also in the focus of the work by Ferraro et al.~\cite{Ferraro2022Measuring}, who propose a measure that reflects to what extent recommendation will familiarize a given user population with certain content categories.

Finally, we mention two examples of works that may lead to a positive and socially desirable \emph{economic} development in quite different ways. First, Pourashraf and Mobasher~\cite{Pourashraf2022Using} target the news recommendation domain and argue that readers' local news interests are different from their preferences for global news.
Thus, by differentiating between local and global news preferences their localized recommendation model increased prediction accuracy with the ultimate goal to revitalize and strengthen local media companies.
The second work from Liu et al.~\cite{Liu2019Personalized} target the domain of \emph{micro-lending}, an approach to provide citizens in impoverished countries with access to capital. In their work, the authors propose a fairness-aware model that aims at ensuring that different demographic groups have a fair chance of being recommended for a micro-loan.

\subsubsection{Environment Protection and Sustainability} 
A diverse set of approaches to build recommender systems `for good' can be found in this category. An early work in this area represents Knijnenburg et al.~\cite{KnijnenburgWB14Smart}, who develop a system to recommend energy-saving measures to users. They specifically explore the impact of different preference elicitation methods on user satisfaction, which in turn influences behavior outcomes. The recommendation of energy-saving measures is also in the focus of the work by Starke et al.~\cite{Starke2021Promoting}. In their work, they analyze the effectiveness of different psychology-informed ways of presenting recommendations to users in a persuasive way.

The reduction of emissions is a central target in other works as well, e.g., in the area of travel and tourism. Bothos et al.~\cite{Bothos2013Choice}, for instance, design a persuasive travel recommender system that nudges users to adopt greener transportation habits. A number of other research works aims at implementing \emph{smart mobility} through recommender systems in the context of \emph{smart cities}, e.g., by promoting the use of public transport and non-motorized mobility options, the reduction of traffic congestions, or addressing parking management problems, see~\cite{QUIJANOSANCHEZ2020SmartCities}. A number of additional recommender systems use cases for smart cities are discussed in~\cite{rafique_developing_2023}, where authors argue that future systems have a higher likelihood of \emph{intent-awareness}~\cite{jannach2024intent}  due to novel technological enablers such as Sensors and Internet-of-Things.

Other recent works in recommender systems focus on \emph{sustainability} in different dimensions. Connected to the previously discussed applications in travel and tourism, Piliponyte et al.~\cite{PiliponyteM023} propose a simulation-based approach to study the effects of personalized advertisement campaigns on tourist choices. A main goal of their work is to support decision-makers in running such campaigns with the goal of avoiding \emph{overtourism}, which may often lead to natural damage. Avoiding overcrowding at certain places is also the focus of the work of Patro et al.~\cite{Patro2020TowardsSafety}.  The goal of their multi-objective recommendation approach is to ensure a better and fairer distribution of businesses, such as restaurants, cafes, or malls, appearing in recommendations, thereby supporting the sustainability of local businesses and the safety of their visitors.

Finally, a larger body of research works targets at promoting environmental protection and sustainability through recommender systems in the \emph{agricultural} domain~\cite{Patel2024MLAgriSurvey}. Various \emph{smart agriculture} use cases can be identified in the literature, including crop recommendation based on various variables related to weather and soil condition, fertilizer recommendation, crop protection and pesticide recommendation or irrigation management. While the literature in this area is rich, we note that most of the proposed recommendation approaches are \emph{non-personalized}. Thus, they are mainly applications of machine learning models for prediction purposes~\cite{Liakos2018ML} or optimization-based solutions for improved decision support.

\subsection{\rsforgood and General Decision Support Systems}
At this point, it is important to iterate what we consider central features of a recommender system and in which ways such systems are different from the class of more general decision-support and advice-giving systems like those just mentioned for the agricultural domain. First, in a recommender system, there commonly is the underlying goal to help users in situations of information overload, where users are confronted with a relatively rich set of options to choose from. Second, we assume that recommender systems are designed to personalize their suggestions based on the needs, preferences and other characteristics of individual users. Third, users are assumed to chose individual items from a given set of \emph{fixed} or preconfigured alternatives, such as picking a movie to watch.

Given this characterisation of the key features of a personalized recommender system, a crop yield prediction system, as for example presented in~\cite{Pande2021Crop}, would not fall under our definition even though called recommender system by the authors. Likewise, there are various other works, e.g., in the health domain, which are sometimes referred to as recommender systems, but are not covered by our definition. The famous MYCIN system~\cite{shortliffe1984rule} from the 1980s, for example, was an expert system implementing hundreds of explicit rules. These rules allowed the system not only to analyze a patient's symptoms, but they could also recommend antibiotics. Nonetheless, we would not consider MYCIN to be a recommender system, as it does not consider the preferences of the physician or the patient. Let us note, however, that the boundary between personalized recommender systems and the more general class of advice-giving systems is sometimes blurred.\footnote{Conversational and knowledge-based recommender systems, for example, typically do not rely on past user preferences. In bundle recommendation scenarios, a recommendation refers to a set of multiple and dynamically combined items. And in future scenarios based on Generative AI, the items may dynamically be created or tailored according to user preferences.}

Certainly, systems like the mentioned agricultural advice-giving solution and medical decision-support systems are important tools that can help promote societal good. Nonetheless, with this essay we primarily aim to address researchers who focus on personalized recommender systems.\footnote{Non-personalized decision-support systems may be covered by more general `AI4Good' initiatives.} We argue that there are plenty of opportunities for such systems also in domains like healthcare and we aim to encourage recommender systems researchers to more often explore the use of personalized solutions in such domains.

\section{\rsforgood: Shifting our Research Focus}
\label{sec:methodology}
Our brief overview in Section~\ref{sec:topics} demonstrates that already a multitude of \rsforgood use cases have been explored in literature. However, the amount of works that focus on \rsforgood-related topics is negligible compared to the extensive body of literature devoted to proposing new algorithms exclusively evaluated offline on the few datasets from the media and e-commerce domains.\footnote{This focus on algorithms and accuracy metrics seems particularly pronounced for central recommender systems publication outlets such as the ACM SIGIR, WWW, KDD and RecSys conference series, as well as for general  AI conferences such as IJCAI or AAAI. Research on \rsforgood use cases, in contrast, does not frequently find its place at these venues.} We re-emphasize here that there is nothing wrong with research in these areas, and that such research may also fall into the scope of \rsforgood. E-commerce recommendations can, for instance, be designed to promote more sustainable products, and a music recommender system's design may target at increasing the exposure of artists from minority groups. Thus, we iterate whether a recommender systems serves a `for good' purpose or not rather depends on the design goals of the system than on the application area. However, not many works in the literature address such `for good' purposes in this way.

The question thus arises why we as a research community do not devote our efforts on research that matters and fosters societal development, instead of continuing to propose increasingly complex machine learning models that we evaluate in terms of their capability to accurately predict held-out data points that derive primarily from the entertainment domain.
And even worse, we do so while knowing that
 various field tests and user studies tell us that increases in offline accuracy may in many cases \emph{not} translate into better value for consumers or businesses~\cite{DBLP:conf/ercimdl/BeelL15,DBLP:journals/tiis/CremonesiGT12,Ekstrand:2014:UPD:2645710.2645737,garcin2014offline,JannachLerche2017,Maksai:2015:POP:2792838.2800184,McNee:2002:RCR:587078.587096}. Given all this evidence for this offline-online gap and many voices from industry, it may actually turn out that much of our research on improving accuracy metrics in offline experiments may be almost entirely worthless in practice.

One part of the answer to the question above may be as simply as this: that it is much easier to do algorithm-oriented research based on offline experiments than to do \rsforgood research. Typical \rsforgood research can have certain ingredients that may be considered challenging. Usually, one needs to understand the idiosyncrasies of the particular use case, commonly leading to a need for interdisciplinary research~\cite{Keestra_Menken_2016}. Moreover, the evaluation of \rsforgood recommender systems can be demanding, as typically a human-centric research approach is required. Moreover, the aspired positive societal impacts can many times only be assessed in a longitudinal perspective. Research based on offline experiments does not face these challenges. Researchers can, for instance, try out various variants of a newly developed algorithm in a largely automated way without the need to involve humans in experiments. However, the predominant use of offline experimentation in combination problematic offline evaluation practices may have led to a certain level of stagnation in our field. To illustrate the predominance of offline experimental designs, we analyzed all papers on recommender systems that were presented at the top-level ACM SIGIR conference between 2022 and 2024. Among the 65 relevant papers, we found no paper that was based on a user study. Furthermore, we could not identify any work that was based on interdisciplinary research. Previous surveys on certain subareas of recommender systems, e.g., \cite{Klimashevskaia2024Survey,jannach2024intent,dejdjoo2023fairness}, underline that user-centric research is largely underrepresented.

\subsection{Offline Evaluation Practices and the Need for a Crisis}
Having said that conducting research based on offline evaluations may be often easier than \rsforgood research does not necessarily imply that it is actually easy to get an algorithm paper published in a top-level outlet. It requires inspiration for a novel technical idea\footnote{In terms of technical proposals, we usually do not observe entirely novel fundamental architectures. More commonly, existing network topologies (e.g., RNN, CNN, GAN, Transformer etc.) and learning approaches (e.g., contrastive learning, multi-task learning) are applied in innovative ways to recommendation problems.}, a working implementation, sometimes a non-trivial mathematical exposition and analysis, as well as in-depth experimental evaluations, which, most recently have become increasingly computationally expensive.

Nevertheless offline evaluation practices might be perceived less effort than \rsforgood research involving humans in the loop. In offline evaluation, the evaluation procedures and metrics are to some extent standardized, whereas in user studies every single aspect of the proposed experimental design may need to be defended against reviewers.\footnote{The heavy bias towards algorithm research is commonly reflected in the composition of the reviewer pools of conferences and journals, and a substantial fraction of the reviewers may not be experts in assessing studies with humans-in-the-loop. They might, for example, dismiss papers on user studies because of an assumedly too low number of participants, being accustomed to working with datasets with thousands of users.}
Furthermore, in offline experiments, there are many degrees of freedom regarding the experimental configuration that is used as a basis to demonstrate an advance with respect to the ``state-of-the-art''. Researchers have some freedom when selecting baselines, evaluation datasets, preprocessing steps, data-splitting techniques, as well as evaluation metrics and even cut-off values for the metrics~\cite{JannachdeSouzaetal2020}. Given this freedom, and knowing about the possible existence of confirmation bias---where the expectation is that the proposed method works well---researchers might unknowingly focus on very specific experimental configurations supporting this bias.

As such, it is not too surprising that a number of  studies in recent years have revealed that the progress that we make in terms of algorithms that lead to better offline accuracy results may actually be quite limited. Quite worrying, these studies show that often decade-old methods or conceptually simple techniques based on nearest-neighbor search can outperform even the latest neural models~\cite{ferraridacremaetal2019,ludewiglatifiumuai2020,ferraridacrema2020tois,RendleMFNeuMF2020,DBLP:conf/recsys/RendleKZK22,shehzad2024performance,RevisitingBPR}.
One main reason for this \emph{phantom progress}~\cite{ferraridacremaetal2019} lies in the apparently common practice of benchmarking a newly proposed and meticulously fine-tuned model against baselines that were not particularly tuned for the given dataset(s). Obviously, nothing meaningful can be concluded from such an experimental setup~\cite{shehzad2023everyone}.
Other reasons may include the use of incorrect or incomplete re-implementations of baseline models, tuning of models on test data, the use of datasets that are not suited for the task, using the same hyperparameters for all compared models `for fair comparison', or reporting the best accuracy values for different metrics across different epochs, see also~\cite{Hidasi2023widespread,Hidasi2023ThirdParty,sun2020we}.

Furthermore, the general level of reproducibility of reported research results turns out to be quite limited as well. While sharing code and data has become more common over the past few years, the majority of published research works do not publicly provide the necessary resources to reproduce the numbers that are reported in the papers~\cite{ferraridacremaetal2019,Cavenaghi2023Repro,jannach2024intent}. This is by no means a novel insight, as similar observations on the reproducibility of information retrieval and recommender systems have been made over the past 15 years \cite{armstrongImprovementsThatDont2009,saidComparativeRecommenderSystem2014}. As the awareness of this problem increases, more and more conferences and journals nowadays provide reproducibility guidelines and encourage authors to share resources. It is nonetheless today still uncommon that papers come with a complete reproducibility package that includes all necessary materials such as the pre-processed datasets, the code of the proposed model and the baselines, information about hyperparameter tuning, detailed documentation and so forth.

Ensuring reproducibility unfortunately however only solves parts of the problems we face with offline experimentation. Let us assume that for a given paper the research methodology is correct and that all resources are shared for reproducibility. We may still need to ask what we can conclude from an offline evaluation that reports a few percentages of accuracy improvements over previously existing models on three \emph{academic} datasets. Given the above-mentioned observations of an offline-online gap, probably not too much. Industry reports from companies like Netflix and Coveo mention the limited predictive power of offline tests, sometimes indicating that practitioners often are ``\emph{shooting in the dark}''~\cite{Gomez-Uribe:2015:NRS:2869770.2843948,rohde2024positionpapershootingdark,Rohde2024Shooting} in only a slightly informed way. Clearly, there are ways in which offline experiments can be helpful. Such experiments might be used to rule out models that perform \emph{very} poorly offline. Furthermore, one can exploit offline tests to analyze certain characteristics of generated recommendations, e.g., in terms of diversity or popularity bias, and then decide which model might suit the given application use case best.

Overall, despite the apparent limitations of offline experiments, still most of the efforts of the community are directed at finding models that lead to small accuracy increases on a limited collection of datasets. Unfortunately, there are no signs of a crisis in the sense of a `turning point' yet~\cite{cremonesi2021aimag}, where we as a community reconsider what kind of problems we address and how we make \emph{contributions that matter} with the goal to impact the real-world in a positive manner.

\subsection{Challenges and Barriers of \rsforgood Research}
\label{subsection:challenges-barriers-rs4good}
\subsubsection{Methodological Challenges}
When reconsidering the \rsforgood use cases discussed in Section~\ref{sec:topics}, we can observe that not many of them can be addressed with traditional offline evaluations. As briefly indicated above, many use cases necessitate a research approach that is driven by the following requirements.

\begin{enumerate}
\item \emph{Interdisciplinary:} Many of the discussed \rsforgood use cases require a deep understanding of the particularities of a given domain. Research in the healthcare domain may be considered a prime example here. A computer science with only layman's understanding of medicine should not be attempting to develop healthcare solutions without the guidance of experts. As a result, research in various \rsforgood domains requires either interdisciplinary collaboration with experts from other domains or at least an in-depth understanding of the idiosyncrasies of the domain.
\item \emph{Human-in-the-Loop and Impact-Oriented Evaluation:} The true value of \rsforgood research for different stakeholders in many cases cannot be assessed through abstract computational measures like precision and recall. We may consider a recommendation-based health intervention or the evaluation of a novel recommendation approach in the e-learning domain as examples. In many of these cases, studies require humans in the loop in the evaluation process and appropriate measures (e.g., degree of health improvement or learning success) that help us assess the impact of the interventions or recommendations on users.
\item \emph{Longitudinal Perspective:} Considering again the examples of health interventions or e-learning applications, the effects of recommendations may only become visible over a longer evaluation horizon, e.g., after several weeks of treatments or at the end of an academic year. Likewise, if we think of using \emph{recommendations as nudges}~\cite{jessejannach2021chbr} towards a desired behavior, e.g., in terms of healthy eating patterns, we must ensure that such behavior changes are sustainable.
\end{enumerate}

An orthogonal problem that seems prevalent in various forms of \rsforgood research is that of understanding, defining, and properly supporting the \emph{goals} that should be achieved through a recommender system, both on an individual and a societal level. To optimally support individual goals like self-actualization described above, a participatory approach seems advisable, where users are asked to explicitly state their goals, as done in~\cite{Liang2024Enabling} for the case of a podcast recommender system. On a more general level, \citet{Ekstrand2015Behaviorism} argue that there are many cases where ``\emph{behaviorism is not enough}.'' While observed user behavior is valuable for recommendation, the authors advocate for increased user involvement and participation in the entire process from algorithm design to evaluation.  Besides the problem of capturing the goals, there might also be trade-off situations to be balanced in goal-setting. Remembering the example from the educational domain described above, it might be tempting to recommend easy courses to students that maximize their short-term success in exams. In the long run, however, a more rounded profile including also challenging materials might be more beneficial both for the student and the society. On the individual level, theory-backed mechanisms for adequate goal setting can be applied to address some of these problems. Several works in the health and personal wellbeing domain for example use the Rasch model and alternative approaches to derive goals and make recommendations that consider both the user's abilities and the difficulty of certain behavioral actions~\cite{Schaefer2019Rasch,Radha2016Lifestyle,Torkamaan2022RecChallenges}.

Certainly, not all \rsforgood use cases may necessarily face all of these
challenges in parallel. For example, there may always be certain types of research without involving humans, for instance, works that are based on simulation approaches like the study in the tourism domain by Piliponyte et al.~\cite{PiliponyteM023} mentioned above. In general, however, research in \rsforgood domains might often be much \emph{slower} and \emph{effortful} in terms of the required research design and the execution of experiments than traditional offline evaluations. Moreover, research on \rsforgood use cases, due to the need to consider domain specifics, may always face the criticism of focusing on a too narrow problem setting, and the generalizability of the findings may be questioned. As a result \rsforgood research may suffer from a basic dilemma of recommender systems research. On the one hand, researchers in computer science generally strive to develop generalizable solutions, e.g., algorithms, that work for a variety of use cases. On the other hand, there is a concrete danger that we overgeneralize and make conclusions regarding the effectiveness of a new algorithm in practice based on a limited set of experiments and  using a specific set of datasets and abstract, domain-agnostic computational measures.

\subsubsection{Research Culture Barriers}
The discussed methodological challenges in most cases will probably imply a slower pace of research and publication. Studies involving human subjects often take significant amounts of time, starting with the time needed to recruit suitable participants. Furthermore, working in interdisciplinary teams, getting approval by ethics boards, understanding domain specifics and collaborating with different institutions (e.g., hospitals and doctors) can be time-consuming and effortful as well. Finally, longitudinal studies like~\cite{Liang2022Exploring,Porcaro2024Assessing} by definition require a longer observation horizon. Such a slower pace of publication however seems largely incompatible with the fast-paced publication culture in computer science.

For the area of machine learning, Turing-award laureate Yoshua Bengio therefore suggests to entirely re-think the publication process, which nowadays is almost focusing on conferences~\cite{bengio2020timetorethink}. This publication process is highly competitive, fast-paced, and driven by deadlines. As a result, Bengio speculates that for many papers that are submitted to conferences and which are eventually published, there might not have been enough time to thoroughly check them for errors. Furthermore, such papers often appear incremental and may lack depth. Ultimately, he advocates for a `slow science'\footnote{\url{https://en.wikipedia.org/wiki/Slow_science}} approach, and envisions a model where research works are rather submitted to fast-turnaround journals, allowing for multiple iterations for improvements, and conference program committees then select already accepted works for presentation and discussion at conferences. This way, the role of conferences can change into a place for connecting with other people and for highlighting and discussing the ``\emph{best and most important ideas}'' within the scientific community.

Following the discussions above, most \rsforgood research will require and benefit from a `slow science' approach. We are aware that the publication process will not rapidly change in the near future. However, we believe that we as community should not shy away from addressing important and from a societal viewpoint relevant research questions just ``\emph{because they are difficult}.''

\section{Summary and Next Steps}
With this opinionated essay, we would like to encourage scholars in the field of recommender systems to more often address problems of societal relevance that matter in practice. Our brief survey of examples of existing \rsforgood research demonstrates that there are plenty of opportunities and open challenges in various areas from healthcare to education and individual and societal development. We however find that successful \rsforgood research requires a major
methodological shift in how we do research, both in terms of our methodology and our publication models.

In terms of the research methodology, more awareness and advanced competencies are required with respect to successfully conducting interdisciplinary research. Various general frameworks for interdisciplinary research have been put forward over the last decades, and several surveys were conducted to understand certain facets, like drivers and barriers, of interdisciplinary collaboration~\cite{Rossini1979Frameworks,Keestra_Menken_2016,Tobi2017MIR,MacRaeFramework2024,siedlok2013organization}. \rsforgood research should build upon these insights from different fields.  Importantly, interdisciplinary research not only means that researchers of different disciplines work together on a problem, but that the work is also based on a shared conceptual model that integrates theoretical frameworks from these disciplines, relies on methodologies that are not limited to one field, and requires the skills of the involved disciplines multiple times throughout the research process~\cite{Aboelela2007Defining}. To further strengthen competencies in such collaborative research endeavours, a stronger emphasis on interdisciplinary research skills in computer science education is needed, as advocated for example in~\cite{Dabu2017CSEducation}.\footnote{Notably, the potential benefits of considering and organizing computer sciences as an interdisciplinary study were discussed decades ago in~\cite{Agresti1976CSInterdisc}.}

Also from a methodological viewpoint, considerable knowledge has been accumulated in different fields regarding appropriate research practices for conducting longitudinal studies. Notable examples can in particular be found in fields such as psychology and medicine~\cite{Schaie1982longi,khoo2006longitudinal}. In~\cite{lynn2009methodology}, methods for longitudinal \emph{surveys} are covered in depth, touching upon a variety of topics like sample design, statistical methods, limitations of longitudinal surveys or ethical questions. Statistical methods for analyzing data that comprise repeated measures for the same individuals over time are covered in depth in~\cite{fitzmaurice2008primer} and in~\cite{Locascio2011Longi}, both works from the medical domain. Overall, various methodologies for conducting longitudinal research have been developed in different fields, which can be applied or adapted for \rsforgood research.

To foster a research culture and publication models that value societal impact, conference organizers could, for instance, consider introducing awards specifically for contributions to societal good. While traditional recognitions, such as Best Paper or Test of Time awards, celebrate technological rigor and lasting relevance in the community, awards for societal impact could highlight research that achieves meaningful and positive change. Such recognitions could play an important role in inspiring the community to pursue impactful \rsforgood problems, aligning the achievements of the community with societal needs.

We are aware that such changes may take their time. In our view, raising awareness on the issues outlined in this essay represents an important first step.
In this context, the organization of topical workshops, seminars, conference tracks and journal special issues may help to increase the attention on \rsforgood topics and refocus future efforts in recommender systems research. Importantly, such venues should be designed to encourage collaboration across disciplines and provide a dedicated space in particular for human-centric works, case studies and longitudinal research.

Generally, we are confident that the community will soon consider personalized ranking algorithms to be a commodity---also due to the increasing capabilities of Large Language Models---and consider the ranking problem to be solved to the extent that a sufficient number of domain-independent well-performing algorithmic solutions are already available by now. Such a development would then allow us as a community to focus on \rsforgood research that has the goal to contribute to a positive development of society.

\bibliographystyle{ACM-Reference-Format}


\end{document}